\documentclass[a4paper,11pt]{article}
\usepackage{pos}

\title{An Euclidean representation of Majorana spins}

\author*[a]{Jacek Wosiek}

\affiliation[a]{Institute of Theoretical Physics, Jagiellonian University,\\
  S. Łojasiewicza 11, 30-348 Kraków, Poland}

\emailAdd{jacek.wosiek@uj.edu.pl}
 
\abstract{An Euclidean representation of bosonized Majorana fermions, prior to imposing constraints, is derived in three space-time dimensons. The difference with the standard three dimensional Ising system is epmhasized. The mild sign problem, does not preclude Monte Carlo simulations in intermediate volumes.
Implementing constraints is briefly outlined.}

\FullConference{%
 The 38th International Symposium on Lattice Field Theory, LATTICE2021
  26th-30th July, 2021
  Zoom/Gather@Massachusetts Institute of Technology
}

\newcommand{\eq}{\begin{equation}}

\newcommand{\eqx}{\end{equation}}
\newcommand{\eqn}{\begin{eqnarray}}

\newcommand{\bi}{\begin{itemize}}

\newcommand{\eqnx}{\end{eqnarray}}
\newcommand{\ei}{\end{itemize}}

\newcommand{\ra}{\rangle}

\newcommand{\nn}{\nonumber}

\begin{document}
\maketitle

\section{Odd lattices, the Hamiltonian and constraints}
In this talk the term "Majorana spins"  refers to a system of spins equivalent to Majorana fermions.
As explained in the previous presentation \cite{BRL,BRW}, a single Majorana fermion can be bosonized  only on lattices with an odd coordination number. The simplest example is provided by a hexagonal lattice in two space dimensions, Fig.1a. It will be convenient to represent it by the equivalent  "brick wall" lattice, c.f. Fig 1b.

\begin{figure}[h]
\begin{center}
\includegraphics[width=5cm]{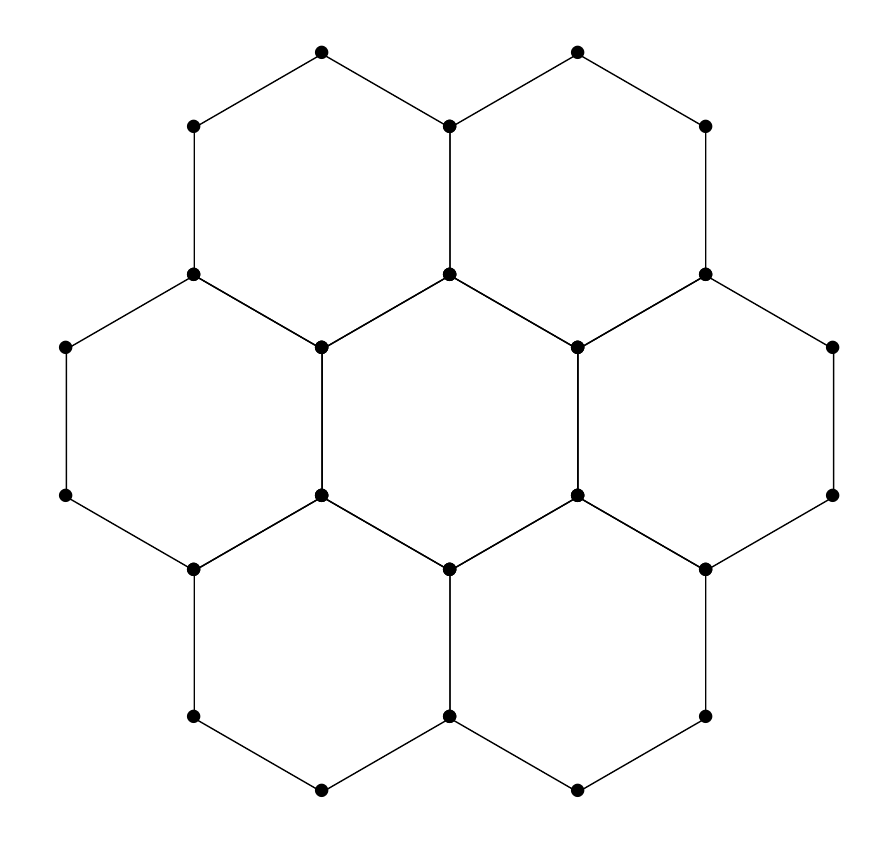}  
\hspace*{3cm} 
\includegraphics[width=6cm]{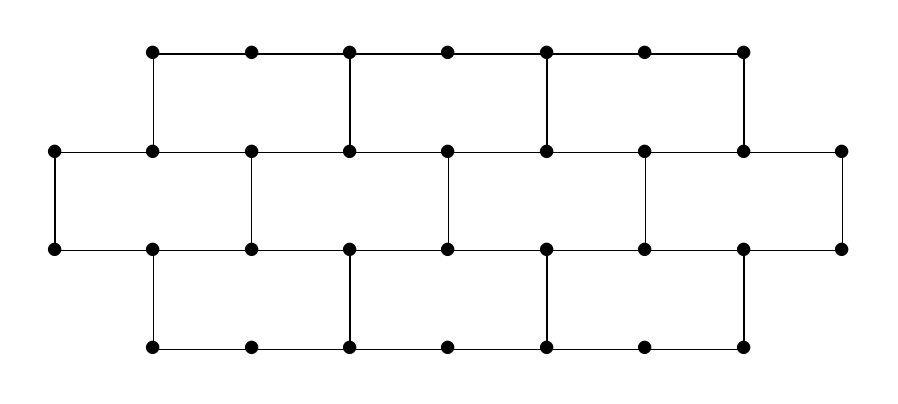} 
\end{center}
\vskip-4mm \caption{A two dimensional lattice with odd coordination number (a) and its rectangular counterpart (b).} \label{fig:f1}
\end{figure}

The equivalent spin hamiltonian reads
\eqn
H&=&H_{kin}+\lambda H_{pot}, \label{H3D}\\
H_{kin}&=&\sum_{l_x} \sigma^x(b_{l_x})\sigma^x(e_{l_x})+\sum_{l_y} \sigma^y(b_{l_y})\sigma^y(e_{l_y}),   \label{kin3}\\
H_{pot}&=&\sum_{l_z} \sigma^z(b_{l_z})\sigma^z(e_{l_z}), \label{hpot}
\eqnx
with links $l_x,l_y$ and $l_z$ organized as in Fig.2, and $b_l$ ($e_l$) standing for the beginning (end) of a link $l$, $\sigma^k$ (k=x,y,z) are the standard Pauli matrices.

\begin{figure}[h]
\begin{center}
\includegraphics[width=6cm]{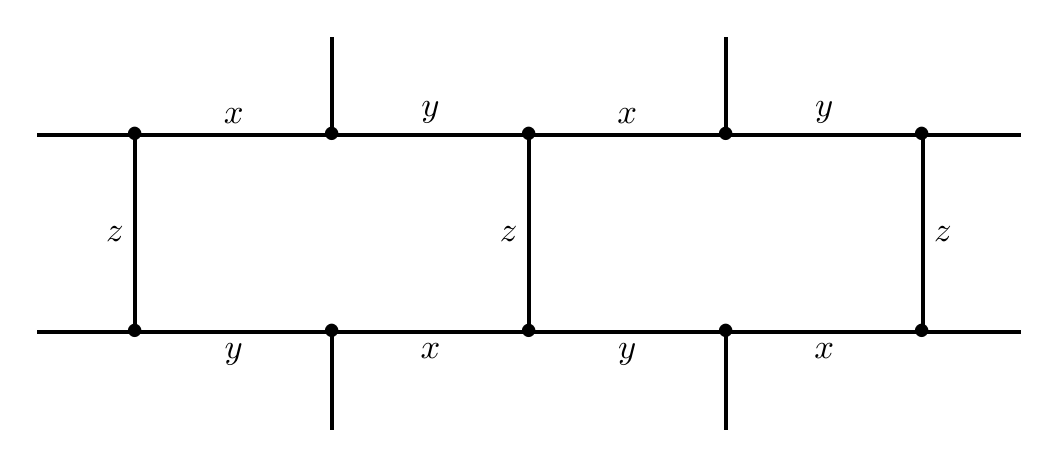} 
\end{center}
\vskip-4mm \caption{Labeling links and corresponding Pauli matrices on a "brick lattice". } \label{fig:f2}
\end{figure}

Together with the local constraints proposed long time ago \cite{W}, and applied to this case in \cite{BRW}, this system is equivalent to a single specie of a free, massless Majorana fermions.

In this contribution an Euclidean representation of the above unconstrained quantum spins will be derived. Implementation of the constraints will be only briefly outlined. More complete discussion in the Hamiltonian formalism can be found in \cite{BRWW}, see also \cite{S,BR2003}.

\section{Euclidean representation of unconstraint
Majorana spins}
Our goal is to construct a Euclidean system of Ising-like spins, with two different couplings, $\beta_t$ and $\beta_s$, which in the continuum time limit  
\eq
\beta_t\rightarrow\infty,\;\;\; \epsilon=e^{-\beta_t}\rightarrow 0,\;\;\; \beta_s=\epsilon\lambda \rightarrow 0, \;\;\;T=1-\epsilon H. \label{ctl}
\eqx
is described by the Hamiltonian (\ref{H3D}).  To this end we follow the procedure of Fradkin and Susskind \cite{FS} and Kogut \cite{K}: $\epsilon$ is an elementary time step and $T$ denotes the transfer matrix of a system.

The time evolution generated by (\ref{H3D}) consists of elementary double-spin flips. This is the main difference with the Ising system, where only a single spin changes during a time lapse $\epsilon$. To better understand this distinction we derive first the euclidean action for the simpler, one dimensional quantum hamiltonian
\eq
H_{1d}=-\sum_n \sigma_n^1\sigma_{n+1}^1 - \lambda \sum_n \sigma_n^3\sigma_{n+1}^3.  \label{H1d}
\eqx

\subsection{Basic idea and the (1+1) dimensional example}
In the Ising model basic trick connecting hamiltonian and euclidean formulations consists of classifying all variations of multiple-spin state into classes with fixed number of single spin flips. 

On the eucliden side, the number of  single flips between two rows is counted by the operator, $\{s\}\equiv s$ and $\{s'\}\equiv s'$,
\eq
S_1(s',s)=\sum_n (s_n - s_n')^2/4\sim -\sum_n s_n s_n',
\eqx
while in the Hamiltonian formulation it is represented by 

\eq
\sum_n \sigma_n^1.
\eqx

In the present case (\ref{H1d}), we are seeking to single out the {\em double spin flips} out of all possible changes of a row 
of spins.
Therefore we begin with the euclidean 8-spin operator which counts {\em disconnected} double flips 
\eqn
S_2^{(8)}&=&\frac{1}{2^4}\sum_n (1+s_{n-1} s_{n-1}')(1-s_{n}s_{n}')(1-s_{n+1} s_{n+1}')(1+s_{n+2} s_{n+2}'). \label{S2V}
\eqnx
Simpler operators can be also used, hence we shall omit the "(8)" superscript if not necessary.

The final, euclidean action should give
in the continuum time limit the lowest (i.e. the leading) weight to above double flips while all other, single and multiple, flips are of higher order in $\epsilon=e^{-\beta_t}$. 

This can be achieved by the following choice for the kinetic part of the action for two rows
\eq
\beta_t L^{kin}(s',s)=\beta_t \left( p(S_1-2 S_2) + S_2 \right). \label{ST}
\eqx
The second term gives the lowest ($=\epsilon$) weight (in the transfer matrix $e^{-L}$) to a single 
 {\em double-flip}. The first one assigns higher (at least $\epsilon^p,  p \geqslant 2$) penalty to all single flips. Excluded from the penalty are two single flips involved in a given double-flip - hence the $(S_1-2 S_2)$ term. 
 
 It is easy to show that the simpler operator
 \eq
S_2^{(6)}=\frac{1}{4^3}\sum_n (s_{n-1} + s_{n-1}')^2(s_{n}- s_{n}')^2(s_{n+1}- s_{n+1}')^2, \label{S1V}
\eqx
also does the job. It prescribes different weights to non-leading transitions but results in the same double flip kinetic part of (\ref{H1d}). This is an explicit illustration of the well known rule that many different euclidean discretizations have the same continuous time limit, hence also the same Hamiltonian. 

Including the potential part is standard and upon convoluting for $N_t$ rows the complete euclidean action for the $N_x \times N_t$ spin follows,

\eq
S= \sum_{x,t}^{N_x,N_t} \left( \beta_t O^{(6)}_{x,t}+\beta_s O^{(2)}_{x,t}\right) \label{Spr1d}.
\eqx
It contains up to six-spin interactions if we choose $S_2=S_2^{(6)}$. Explicitly 
\eq
O^{(6)}_{x,t}=\left( \frac{1-2 p}{4^3} (s_{x-1,t} + s_{x-1,t+1})^2(s_{x,t}- s_{x,t+1})^2(s_{x+1,t}- s_{x+1,t+1})^2
\right)
\eqx
\eq
+\frac{p}{8} (s_{x,t} - s_{x,t+1})^2+\frac{p}{8} (s_{x+1,t} - s_{x+1,t+1})^2,
\eqx
and
\eq
O^{(2)}_{x,t}= - s_{x,t}s_{x+1,t}.
\eqx
The integer, two dimensional lattice coordinates are now labelled as $(x,t)$.

This concludes our construction of the two dimensional, euclidean system which in the continuum time limit is described by the Hamiltonian (\ref{H1d}).

There is no obvious symmetry between space and time directions. That is why $\beta_s$ and $\beta_t$ are left different. One could relate them, e.g. $4\beta_t=\beta_s$, but this will not restore the exact symmetry in any obvious way. 
One might attempt to restore the full space-time symmetry in the continuum limit by tuning both couplings such that the correlation lengths in both directions are the same.

\subsubsection{$\sigma^2\sigma^2$ terms - the phases}
The second example deals with the phase generating kinetic terms
\eq
H_{1d}^{ph}=-\sum_{n\; even}\sigma_n^1\sigma_{n+1}^1 -\sum_{n\; odd}\sigma_n^2\sigma_{n+1}^2 - \lambda \sum_n \sigma_n^3\sigma_{n+1}^3,  \label{H12}
\eqx
still in one space dimension.

 Begin with an evolution of a two spin system $s=\{s_1,s_2\}\rightarrow s'=\{s_1',s_2'\}$. As far as the change of spin states is considered, the action of $\sigma^2\sigma^2$ is the same as that of 
 $\sigma^1\sigma^1$. The only difference is the proportionality factor between the two results
 \eq
 \sigma^2\sigma^2 | s_1,s_2 \ra = \eta  \sigma^1\sigma^1 | s_1,s_2 \ra = \exp{\left(  \frac{i \pi}{2}(s_1+s_2)\right)}  \sigma^1\sigma^1 | s_1,s_2 \ra.\label{phase}
 \eqx
 Generalization to the whole row of spins is straightforward. The kinetic term of the hamiltonian (\ref{H12}) will be reproduced by the action (\ref{ST}) supplemented by a phase
 (\ref{phase}) for each odd link. This gives for the new action of the two complete rows (and with the unchanged diagonal potential term)
\eqn
L_{ph}(s',s)&=&\beta_t  \left(2(S_1-2 S_2) + S_2 \right)+ \beta_s S^{pot} \nn \\
&+&\frac{i \pi}{2} \frac{1}{2^3}\sum_{x - odd} (s_x+s_{x+1}) (1+s_{x-1} s_{x-1}')(1-s_{x}s_{x}')(1-s_{x+1}s_{x+1}'). \label{L2ph}
\eqnx  
  Iteration of the corresponding transfer matrix does not introduce any new elements. The final action  of the two dimensional system reads

\eqn
S_{2D} &=& \sum_{x,t} \left( \beta_t O^{(6)}_{x,t}+\beta_s O^{(2)}_{x,t}\right) +  \frac{i \pi}{2}\sum_{x-odd,t} O^{(7)}_{x,t},  \label{Sph} \\
O^{(7)}_{x,t} &=& \frac{1}{2^3}(s_{x,t}+s_{x+1,t}) (1+s_{x-1,t} s_{x-1,t+1})(1-s_{x,t}s_{x,t+1})(1-s_{x+1,t}s_{x+1,t+1}).  \nn
\eqnx
  In words: only double flips of pairs of spins, sitting on odd links, generate the euclidean phase. This is in accord with the Hamiltonian (\ref{H12}).

\subsection{2+1 dimensional system}
Derivation of the euclidean action of a three-dimensional $(x,y,t)$, periodic in all directions, system is very similar to the previous (1+1) example. 

The two kinetic terms (\ref{kin3})  are represented by the same six- or eight-spin couplings between the adjacent time slices plus the appropriate phase, which naturally generalizes 
the (1+1) dimensional phase (\ref{Sph}) to three euclidean dimensions.

On the other hand diagonal, in the hamiltonian form, potential terms (\ref{hpot}) are represented by the standard Ising-like, ferromagnetic couplings along the y-direction. They are located on the shorter edges of bricks at each time slice. Hence, they are staggered in  accord with the (t-independent) x-y parity, $\zeta_{xy}=(-1)^{x+y}$, of a site originating given $l_z$-link in (\ref{hpot}). The final action reads
\eq
S_{3D}= \beta_t \sum_{x,y,t} O^{(6)}_{x,y,t}+\beta_s \sum_{x,y,t,\zeta_{xy}=1} O^{(2)}_{x,y,t} +  \frac{i \pi}{2}\sum_{x,y,t,\zeta_{xy}=-1}O^{(7)}_{x,y,t}, \label{S3D}
\eqx
with the phase operator $O^{(7)}_{x,y,t}$ being the direct generalization of  above $O^{(7)}_{x,t}$ to three dimensions
\eq
O^{(7)}_{x,y,t} = \frac{1}{2^3}(s_{x,y,t}+s_{x+1,y,t}) (1+s_{x-1,y,t} s_{x-1,y,t+1})(1-s_{x,y,t}s_{x,y,t+1})(1-s_{x+1,y,t}s_{x+1,y,t+1}).
\eqx
Similarly for other couplings
\eqn
O^{(6)}_{x,y,t}&=& \frac{1-2 p}{8} (1+s_{x-1,y,t} s_{x-1,y,t+1})(1- s_{x,y,t} s_{x,y,t+1})(1- s_{x+1,y,t} s_{x+1,y,t+1})
+ \frac{p}{2}(1- s_{x,y,t} s_{x,y,t+1}) \nn\\
O^{(2)}_{x,y,t}&=& - s_{x,y,t}s_{x,y+1,t}.
\eqnx
The action (\ref{S3D}) describes then a three dimensional Ising-like system. Together with constraints, described in the accompanying talk, it provides an equivalent, euclidean representation of a single, quantum Majorana spin on a two dimensional, spatial lattice. 

Even without constraints, the Boltzmann factor associated with  (\ref{S3D}) is not positive. However the origin of its phases is now conceptually simple. Below we look how severe is the sign problem in these unconstraint, euclidean models.

\section{The sign problem}

The standard way to deal with non-positive weights consists of the reweighting \cite{LB}. Instead of potentially  negative Boltzmann factor $\rho=\exp{(-S)}$, one uses as a MC weight its absolute value $\rho_A=|\rho|$, correcting at the same time all observables for this bias.

The sign 
of the exact Boltzmann factor 
\eq
<sign>\equiv<\frac{\rho}{\rho_A}>_A= \frac{Z}{Z_A},
\eqx
averaged over the modulus $\rho_A$, gives us some idea how practical is the trick. If this average is close to 0 the method fails. On the other hand, even small but non-vanishing, at large volumes, values of $<sign>$ allow to expect meaningful estimates.

We have calculated analytically  above average for both (1+1) and (2+1) dimensional models by employing the transfer matrix  technique for a range of small volumes. It is seen below that the sign problem does not seem to be very severe in this circumstances. Consequently MC studies remain a viable approach to explore these systems in detail, at least for the intermediate volumes.

\begin{figure}[h]
\begin{center}
\includegraphics[width=13cm]{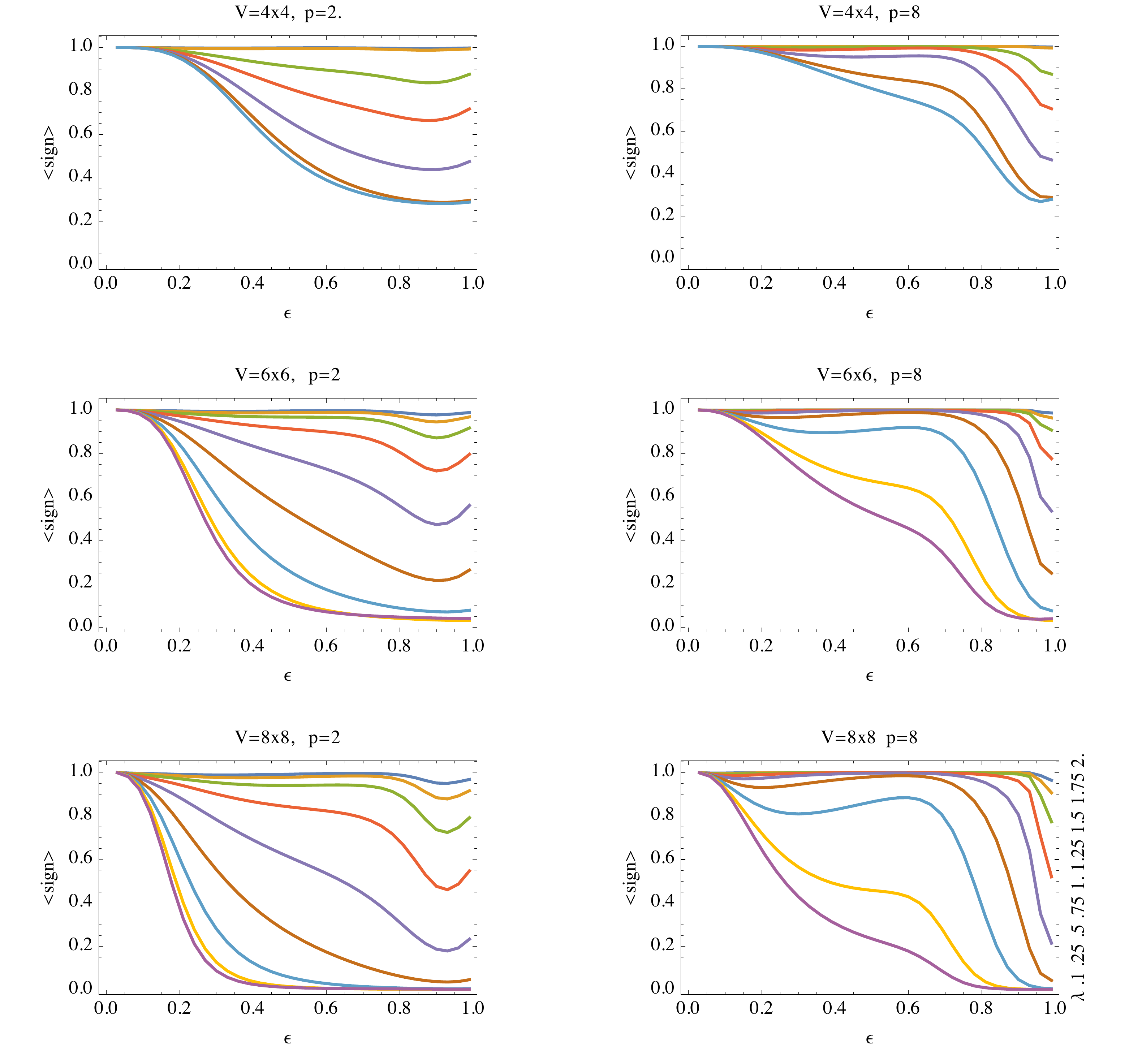}
\end{center}
\caption{Exact results for the average sign as described in the text. Assignment of different $\lambda$ values is defined in the last panel.} \label{fig:f3}
\end{figure}

\subsection{(1+1) dimensions}
Partition functions $Z$ and $Z_A$ were calculated exactly by summing Boltzmann factors $\exp{(-S_{2D})}$ and $|\exp{(-S_{2D})}|$, as defined in (\ref{Sph}).
 The average sign is shown in Fig.\ref{fig:f3}, for a range of two dimensional volumes and various couplings $\beta_t$ and $\beta_s$. To see clearly the continuum time limit, the results are displayed as a function of a time step, $\epsilon=\exp{(-\beta_t)}$, and parametrized by different couplings $\lambda=\beta_s/\epsilon$ in the hamiltonian (\ref{H12}). Second column displays analogous results for larger penalty parameter $p$.

Indeed the sign problem seems manageable for a sizeable part of the parameter space. In the continuum time limit it vanishes entirely. 

Increasing the penalty parameter, $p$, also helps since then undesired transitions vanish faster with $\epsilon$.

Both of these features seem to be universal, i.e. they show up also in our three dimensional system. They can be readily understood and used for our advantage, as discussed below.

\begin{figure}[h]
\begin{center}
\includegraphics[width=10cm]{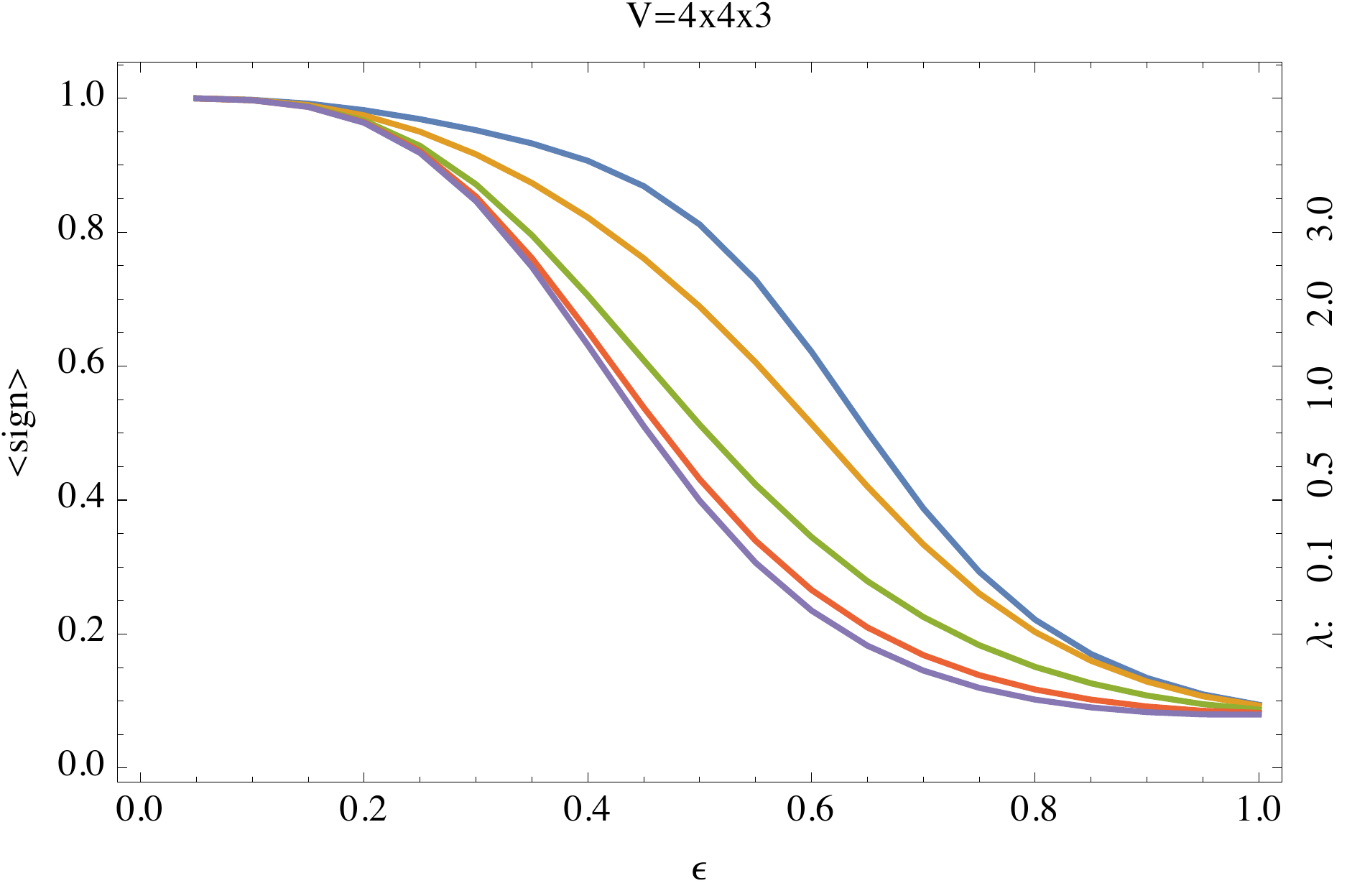}
\includegraphics[width=10cm]{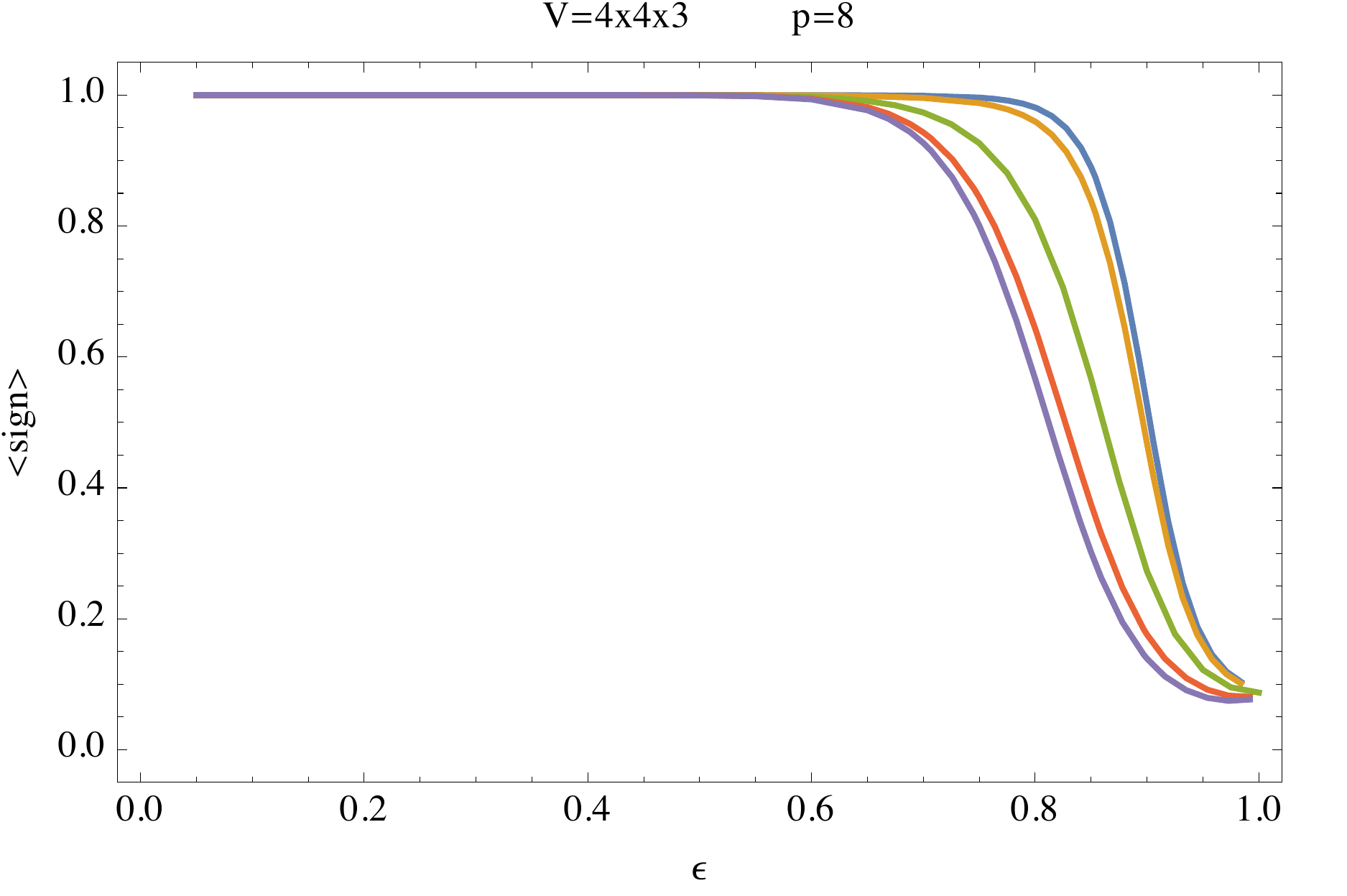}
\end{center}
\caption{Average sign in the three dimensional case.} 
\label{fig:f4}
\end{figure}

\subsection{(2+1) dimensions}
 For the three dimensional euclidean system, (\ref{S3D}) a brute force summation of all $2^V$ terms becomes already a challenge. Still it was possible to obtain $<sign>$ for $V=L_x L_y L_t=4\times 4\times 3$, as shown in Fig.\ref{fig:f4}. It was done by constructing two subsequent transfer matrices in the $y$ direction. 

Again, as in the (1+1) dimensions, the phase is harmless for small $\epsilon$ and this feature improves dramatically with increasing the penalty parameter.

In addition for $L_t=2$ no phase was observed in all cases. That is $<sign>=1$ for all values of parameters and for all studied dimensions.

\section{The summary and outlook}
 
All the regularities observed above can be readily understood and generalized for arbitrary sizes of lattices, providing at the same time some guidelines for other, similar systems.

Consider first the case of $L_t=2$. The partition function 
\eq
Z^{(2)}=Tr T^2 = \sum_{I,J} T_{I,J} T_{J,I},
\eqx
is the sum over two composite states of spins at the two time slices. The non-zero phase can occur only if I and J differ by a double flip. However in this case the phases of $T_{I,J}$ and $T_{J,I}$ cancel and the result is positive for each pair of configurations.

On the other hand, already for $L_t=3$ there are three states in the game
\eq
Z^{(3)}=Tr T^3 = \sum_{I,J,K} T_{I,J} T_{J,K} T_{K,I}.
\eqx
Hence a single double-flip, e.g. in I$\rightarrow$J, can be balanced by two subsequent single-flips in J$\rightarrow$K and K$\rightarrow$I transitions. Since a phase occurs only in the double flip transition I$\rightarrow$J, this particular contribution will be negative and would contribute to $<sign>  < 1$.

Consequently, the single flip transitions provide an undesired background which indirectly causes negative signs of Boltzmann factors, hence the sign problem. 

However such transitions vanish in the continuum time limit having a weight of the higher order in $\epsilon$. This is clearly confirmed by our calculations, c.f. Fig.\ref{fig:f3} and Fig.\ref{fig:f4}, and explains why in the continuum time limit the sign problem vanishes. 

Moreover, by increasing the penalty parameter $p$ we can force the "bad transitions"  to vanish faster. Indeed this is also confirmed by our results for $p=8$ in both dimensions. 

An attractive possibility is to set $p=\infty$. This should still leave us with the local action with a new type of local constraints. In such a system  negative weights would not be allowed {\em at all}. 

By reversing this logic one might  implement the constraints, required by the exact bosonization, in a form of a new plaquette coupling with an "euclidean Lagrange multiplier", $\mu$, say and perform simulations with a finite, but sufficiently large $\mu$.

We are looking forward to study all these options more quantitatively in the furture.

\section*{Acknowledgements}

The talk based on the common work with Arkadiusz Bochniak and B\l a\.zej Ruba \cite{BRW}. This work is supported in part by the NCN grant: UMO-2016/21/B/ST2/01492.


\end{document}